\documentstyle[12pt]{article}
\advance\textheight by \topskip
\begin{document}

\rightline{SU-ITP-92-29}
\rightline{hep-th/9211015}
\rightline{\today}

\vspace{.8cm}
\begin{center}
{\large\bf ENTROPY AND ACTION OF DILATON BLACK HOLES \\}

\vskip .9 cm

{\bf Renata Kallosh \footnote {On leave  from: Lebedev
Physical Institute, Moscow. \  E-mail address:
kallosh@physics.stanford.edu},
Tom\'as Ort\'\i n \footnote{Bitnet address: tomaso@slacvm} and Amanda
Peet
\footnote{E-mail address: peet@slacvm.slac.stanford.edu} }
 \vskip 0.1cm
Physics Department, Stanford University,
Stanford    CA 94305
\end{center}
\vskip .6 cm
\centerline{\bf ABSTRACT}
\vspace{-0.7cm}
\begin{quotation}

We present a detailed calculation of the entropy and action of
$U(1)^2$ dilaton
black holes, and show that both quantities coincide with one quarter
of the
area
of the event horizon. \vskip 0.2 cm

 Our methods of calculation make it possible to find an  explanation
of
the rule $S = A/4$ for all static, spherically symmetric
black holes studied so far. We show that the only
contribution to the entropy comes from the extrinsic curvature term
at the
horizon, which gives $S = A/4$ independently of the charge(s) of the
black
hole,
presence of scalar fields, etc.  Previously, this result did not have
a
general explanation, but was established on
a case-by-case  basis.  \vskip 0.2 cm

The on-shell Lagrangian for maximally
supersymmetric extreme dilaton black holes is also calculated and
shown to
vanish, in agreement with the result obtained by taking the limit of
the
expression obtained for black holes with regular horizon. \vskip 0.2
cm

The physical meaning of the entropy is discussed in relation to the
issue of
splitting  of extreme black holes.

\end{quotation}

\normalsize
\newpage

\section{Introduction }

 It has been known  for some time that the area of the event horizon
of a black
hole behaves like the entropy of a thermodynamic system
\cite{Bekenstein},\cite{BCH}. After Hawking found that  the
temperature $T$ of the black hole thermal radiation was
related to the surface gravity $\kappa$ as $T = \frac{\kappa}{2\pi}$,
it became
clear that the analogy could be made more precise by identifying the
entropy
with one quarter of the area of the horizon. However, the physical
origin of
this
identification was obscure.

Gibbons and Hawking gave more support to this
identification, by  performing a direct calculation of the partition
function
in
the saddle point approximation, obtaining the same result for the
black holes
known at that time (the Kerr-Newmann family) \cite{GH}.  A more
recent
calculation for rotating black holes may be found in \cite{York}.
The result
is to
be interpreted as the intrinsic entropy of the gravitational field,
even in the
absence of thermal gravitons.

However, a general proof of the relation $S = A/4$ was absent.  Each
time when
a new
class of black holes was discussed, it was necessary to perform
rather
complicated calculations anew, and, surprisingly enough, all results
obtained
so far always supported the simple rule $S = A/4$. In particular, in
one of our
previous papers we presented the results of our
calculation of the action/entropy for the recently discovered family
of
stringy dilaton
black holes. We again found that the entropy is given by one quarter
of
the area of the event horizon \cite{US}.   The purpose of this paper
is
two-fold.
First of all, we present all the explicit calculations for the
stringy dilaton
black
holes, making a careful distinction between the entropy and the
action. On the
other hand, we are giving a general explanation of the rule $S =
A/4$, which is
applicable to all static, spherically symmetric
black holes studied so far.  We will use the conventions given in
ref.
\cite{US}.

In sec. 2 we calculate the entropy
of a general static spherically symmetric black
holes. We generalize Hawking's treatment of Schwarzschild
black hole \cite{Hawk100} and see that there is
no need to calculate the action to get the entropy. In fact, one has
to
calculate
only the contribution of the surface term (the integral of the
extrinsic
curvature
$K$) on the horizon. All the gauge terms are shown to drop out.
We then calculate this term for general static spherically symmetric
black
holes with regular horizon, and find that it always gives us one
quarter of the
area of the horizon.

In sec. 3 we calculate the Euclidean action for  $U(1)^2$
dilaton black holes. First we show that the on-shell bosonic action
of
dimensionally reduced string theory is a total derivative.  Even
though
the gauge terms do contribute and the extrinsic curvature surface
term is
calculated at infinity, we get the same result: action equals
entropy.

In sec. 4 we calculate the Lagrangian for the $N=2$ supersymmetric
black holes
directly as the integrand of the volume integral, keeping all total
derivative
terms. This calculation explains, from the point of view of restored
$O(2)$
symmetry between the two central charges, why the entropy of the
maximally
supersymmetric black holes vanishes.

The last section contains some discussion of the puzzle surrounding
the
physical
meaning of the entropy of extreme dilaton black holes and a possible
relation
to
the issue of splitting of them.

 \section{Entropy}

The starting point in the thermodynamic study of a statistical system
is the
calculation of a thermodynamic function or potential. In the presence
of a set
of conserved charges $C_{i}$ and their related potentials $\mu_{i}$
it is
convenient to work in the grand canonical ensemble, where the
fundamental
object is the grand partition function
\begin{equation}
{\cal Z}=Tr e^{-\beta (H-\mu_{i}C_{i})} \ ,
\end{equation}
and the thermodynamic potential
\begin{equation}
W=E-TS-\mu_{i}C_{i} \ ,
\end{equation}
is related to the grand partition function by
\begin{equation}
e^{-\beta W}={\cal Z} \ .
\end{equation}

Once  ${\cal Z}$ is known, all thermodynamic properties of the system
can
be obtained; for example, the entropy is given by
\begin{equation}
S  =  \beta (E-\mu_{i}C_{i})+\log {\cal Z}\ .  \label{definition}
\end{equation}

Gibbons and Hawking \cite{GH} discovered that the Euclidean
partition function of quantum gravity, when evaluated in the saddle
point
approximation expanded about one of the black hole metrics known at
that
time, can be interpreted as an approximation to the {\it thermal}
grand
partition
function of a system of temperature equal to the black hole
temperature.
In this semiclassical approximation, and from the path integral
representation
of the partition function, we have
\begin{equation}
{\cal Z}= e^{-I_{\phantom{}_{E}}}\ ,
\end{equation}
where $I_{\phantom{}_{E}}$ is the Euclidean on-shell action. An
important
observation made in ref. \cite{GH} is that the Euclidean sections of
the
complexified metrics studied have only one boundary, spatial infinity
$r\rightarrow \infty$. The reason is that the region inside the
horizon is not
present and that the manifold is regular on the horizon, provided
that the
Euclidean time $\tau$ has period $\beta=T^{-1}$. The extrinsic
curvature
surface term present in $I_{\phantom{}_{E}}$ has to be calculated
only at
infinity. We stress
this fact by using the notation $I_{\phantom{}_{E}}^{\infty}$,
\begin{equation}
\log{{\cal Z}}=-I_{\phantom{}_{E}}^{\phantom{}^\infty}\ .
\label{logZ}
\end{equation}

Moreover, it was found that the action of the black hole
metric was equal to the entropy calculated using  eq.
(\ref{definition}), and
both were equal to one quarter of the area of the event horizon, the
value
suggested by the first law of black hole thermodynamics \cite{BCH}.

The fact that the action coincides with the entropy was explained by
Hawking
using scaling arguments in ref. \cite{Hawk100}. In the same
reference, he also
gave a prescription to calculate independently the term $\beta E$ in
the same
approximation for the Schwarzschild case in which there are no
charges
involved.
To calculate the (mean value of the) energy one has to calculate the
action by
taking into account also the contribution to the surface
term from the horizon. The $\tau=constant$ surfaces have two
boundaries, at
the horizon and at infinity. In  the study of imaginary time
evolution between
two surfaces $\tau_{1}$ and $\tau_{2}$ with these given boundary
conditions,
the dominant contribution to the path integral comes from the
Schwarzschild
metric, and both boundaries have to be taken into account. One has
\begin{equation}
<\tau_{1}|\tau_{2}>=e^{-(\tau_{2}-\tau_{1})E}\ ,
\end{equation}
and when $\tau_{2}-\tau_{1}=\beta$,
\begin{equation}
\beta E=I_{\phantom{}_{E,h}}^{\phantom{}^\infty}\ .
\end{equation}

Again, the notation $I_{\phantom{}_{E,h}}^{\phantom{}^\infty}$
stresses the
fact that the contribution of
the horizon has to be taken into account in the surface term. In the
presence of conserved charges, one can consider constrained imaginary
time
evolution so that only metrics with the prescribed charges are
considered in
the
path integral. This can be implemented by using Lagrange multipliers
$\mu_{i}$. The analogues of the above expressions are
\begin{equation}
<\tau_{1}|\tau_{2}>=e^{-(\tau_{2}-\tau_{1})(E-\mu_{i}C_{i})}\ ,
\end{equation}
and when $\tau_{2}-\tau_{1}=\beta$,
\begin{equation}
\beta(E-\mu_{i}C_{i})=I_{\phantom{}_{E,h}}^{\phantom{}^\infty}\ .
\label{betaE}
\end{equation}

Now, if we substitute eqs. (\ref{logZ}), (\ref{betaE}) into eq.
(\ref{definition}) we get
\begin{equation}\label{ieiehi}
S
=I_{\phantom{}_{E,h}}^{\phantom{}^\infty}-
I_{\phantom{}_{E}}^{\phantom{}^\infty}\ .
\end{equation}
These two terms are explicitly given by
\begin{eqnarray}\label{ieie}
I_{\phantom{}_{E,h}}^{\phantom{}^\infty} & = &
\frac{1}{16\pi}\int_{M}(-R+
{\cal L}_{matter}) +
\frac{1}{8\pi}\int_h^\infty [K]  \, \ , \nonumber \\
\nonumber\\
 I_{\phantom{}_{E}}^{\phantom{}^\infty} & =  &
\frac{1}{16\pi}\int_{M}( -R +
{\cal L}_{matter}) +
\frac{1}{8\pi}\int^\infty [K] \, \ , \end{eqnarray}
and finally, substituting eqs. (\ref{ieie}) into eq. (\ref{ieiehi}),
we have
for
the entropy of a general black hole

\begin{equation}\label{generalentropy}
S  = \frac{1}{8\pi}\int_h  [K]\ .
\end{equation}

\noindent i.e. simply the {\it extrinsic curvature surface term at
the
horizon}.
This is a remarkable fact that emphasizes the intrinsic gravitational
nature
of the entropy so calculated.

The next step is to calculate $[K]$ for a sufficiently general case.
For us it
will
be general enough the case of a static, spherically symmetric
asymptotically flat black hole. Extreme purely electric and magnetic
black
holes have no regular horizon and we will treat them in the last
section.
 Let us start with the definition of $K$, the trace of the extrinsic
curvature
of
the hypersurface with (spacelike) unit normal vector $n^{\mu}$
\begin{equation}
K = h^{\mu \lambda} \nabla_\mu n_\lambda \ ,
\end{equation}
where
\begin{equation}
h_{\mu\nu} = g _{\mu\nu}+n_\mu n_\nu
\end{equation}
is the induced metric on the hypersurface.  The term $[K]$ in
eq. (\ref{generalentropy}) is the difference $K-K_{0}$, where $K_{0}$
is
obtained by substituting into $K$ the flat space metric. An infinite
contribution
is subtracted in this way, so that we obtain finite results for the
action.
However,
were the results obtained without the subtraction of $K_{0}$ finite,
it would
not
be needed.

For a spherically symmetric metric
\begin{equation}\label{sphmet}
ds^2 = g_{tt} dt^2 + g_{rr} dr^2 - r^2 d\Omega^2 \ ,
\end{equation}
we find that
\begin{equation}
K =  \frac{1}{\sqrt{-g_{rr}}} \Biggl[ \frac{1}{2}  \frac{\partial_r
g_{tt}}{g_{tt}} + \frac{2}{r} \Biggr] \qquad {\mbox{and}} \qquad
 K_0 =  \frac{2}{r}\ ,
\end{equation}
for surfaces of constant $r$.
Recalling that the Euclidean time $\tau$ is compactified on
$[0,\beta]$, we
obtain for the  surface integral calculated with outward-pointing
$n^{\mu}$

\begin{equation}\label{Katr}
\frac{1}{8 \pi} \int_r{[ K ]} = -\frac{1}{\kappa}
\Biggl(\frac{A(r)}{4} \Biggr) \Biggl[ \frac{1}{2} \frac{\partial_r
g_{tt}}{\sqrt{-g_{rr} g_{tt}}} + \frac{2}{r}
\sqrt{\frac{g_{tt}}{-g_{rr}}} -
\frac{2}{r} \sqrt{g_{tt}} \Biggr]
\end{equation}

\noindent where the surface gravity
$\kappa = \frac{2\pi}{\beta}$ is given by
\begin{equation}\label{surgrav}
\kappa = \frac{1}{2} { \frac{\partial_r g_{tt}}{\sqrt{-g_{rr}
g_{tt}}} }
\biggm|_{r=r_h} \, .
\end{equation}
If the horizon  occurs at a finite value of $r$, which we denote by
$r_{h}$,
and
$g_{tt}=0={{g_{tt}}\over{g_{rr}}}$,  we see that  the term
$\frac{1}{\kappa}$
coming from the $\tau$ integration is cancelled by the first term in
the square
brackets by using expression (\ref{surgrav}) for the surface gravity,
yielding

\begin{equation}\label{Kobs}
S= -\frac{1}{8\pi}\int_{h}
[K]  = \frac{A(r_h)}{4}\ .
\end{equation}

\noindent Thus we have found that the entropy is again one quarter of
the area
of the event horizon. This result has been established for the
general class of
static spherically symmetric black holes (in particular, this
includes
charged axion-dilaton black holes) with metric given in eq.
(\ref{sphmet}).

 \section{Action}

\subsection{The on-shell action for axion-dilaton black holes is
topological}

The total (bosonic) action for stringy $d=4$ dilaton-axion black
holes is given
by the volume integral
 \begin{equation} \label{action}
 I = \frac{1}{16\pi}\int d^4x\, \sqrt{-g}\, \left( {\cal L}_{grav} +
{\cal L}_{dil} + {\cal L}_{axion} +{\cal L}_{ gauge} \right) \ .
\end{equation}

We will follow the presentation of the gravitational
part of the action as given in \cite{LL}, but using the notation of
\cite{US}.
\begin{equation}\label{LL}
\sqrt{-g}\,{\cal L}_{grav} =  \sqrt{-g} (-R) +
\partial_\mu  \sqrt{-g}\, \omega^\mu \ ,
 \end{equation}
where the vector $\omega^\mu$ in the total derivative term in the
gravitational Lagrangian is
\begin{equation}\label{eqforomega}
\omega^\mu = g^{\lambda \rho} \Gamma^{\mu}_{\lambda \rho} -
g^{\lambda \mu} \Gamma^{\nu}_{\lambda \nu}\ .
\end{equation}

In general, an infinite contribution (like the $K_{0}$ term of the
previous
section) will have to be subtracted from eq. (\ref{action}) in order
to obtain
finite results.

In the $SO(4)$ version of the action of $N=4, d=4$
supergravity, or dimensionally reduced string theory,
\begin{eqnarray}
{\cal L}_{dil}      & = & 2\partial^\mu \phi \cdot \partial_\mu \phi
\ ,\\
{\cal L}_{axion}  & = & \frac{1}{2}
{\mbox{e}}^ {4\phi} \partial^\mu a \cdot \partial_\mu a   \ ,
\\
{\cal L}_{ gauge} & = & - \left(
{\mbox{e}}^{-2\phi}F_{\mu\nu}^n F^{n\,  \mu\nu} + {\mbox{e}}^
{2\phi}\tilde
G_{\mu\nu}^n\tilde G^{n \, \mu\nu}\right)
+\nonumber \\
& &+ \,  i \,  a \left(
{ F_{\mu\nu}^n * F^{n\, \mu\nu} + \tilde G_{\mu\nu}^n
* \tilde G^{n\,\mu\nu} } \right) \ ,  \label{a}
\end{eqnarray}
where $g_{\mu\nu},\phi, a, A_\mu ^n, \tilde B_\mu ^n , n=1,2,3$,
are the metric, dilaton, axion, and six vector fields.  In order to
calculate
the
on-shell action we will need the following equations of motion:
\begin{eqnarray}\label{R}
-R  +2 \nabla^{\mu}\phi\cdot \nabla_{\mu}\phi  + \frac{1}{2}
{\mbox{e}}^ {4\phi}\partial^\mu a \cdot \partial_\mu a & = &0\ ,  \\
\nabla_\mu({\mbox{e}}^{-2\phi} F^{n\, \mu\nu} - i a * F^{n\, \mu\nu})
=
\partial_\mu \left (\sqrt{-g} ({\mbox{e}}^{-2\phi} F^{n\, \mu\nu}
- i a * F^{n\, \mu\nu}) \right ) &=& 0  \label{eqmo} \, ,
\end{eqnarray}
and the corresponding equations for $\tilde G^{n\, \mu\nu}$.

If we use eqs. (\ref{R}), (\ref{eqmo}) in the action (\ref{action}),
we end up
with
the following on-shell action,
which is a total derivative:
\begin{equation}\label{useful}
I_{on-shell} = \frac{1}{16\pi} \int d^4x\ \partial_\mu
\Biggl[\sqrt{-g}\,
\left( \omega^\mu + A_\nu^n ({\mbox{e}}^{-2\phi} F^{n\, \mu\nu} -
i a * F^{n\, \mu\nu})+ \tilde B_\nu^n ({\mbox{e}}^{+2\phi}
\tilde G^{n\, \mu\nu} -  i a * \tilde G^{n\, \mu\nu}) \right) \Biggr]
\ .
\end{equation}

An equivalent form of the on-shell action can be given in terms of
differential forms.   The gauge part of the on-shell action was
calculated for
Einstein-Maxwell theory in \cite{BrillRB} and found to be an exact
differential
form; we generalize that procedure here.  In addition, we will need
the
gravitation action as a form.  For this purpose, one may start with
the
gravitational action in terms of tetrad and spin connection forms
\begin{equation}\label{form} \int \sqrt{-g}\,{\cal L}_{grav} = \int
e_a\wedge
e_b\wedge * R^{ab} -  d \, (e_a\wedge e_b\wedge * \omega^{ab})\ .
 \end{equation}
The on-shell action takes the following form:
\begin{equation}
I_{on-shell} = \frac{1}{16\pi} Tr \int d \, \biggl[ * \omega +
A\wedge (e^{-2\phi}  *F-iaF )  +
\tilde B \wedge (e^{+2\phi}  *\tilde G -ia\tilde G )\biggr] \ ,
\label{4form}\end{equation}
where $Tr$ on the vector fields means sum over all vector fields and
\begin{equation}
Tr * \omega = (e_a\wedge e_b\wedge * \omega^{ab}) \ .
\end{equation}

Thus we have shown that the on-shell bosonic part of the $SO(4)$
supergravity
action is an integral over the exact differential form $d \Xi$, where
\begin{equation}
\Xi = \frac{1}{16\pi} {\mbox {Tr}} \biggl[* \omega + A\wedge
(e^{-2\phi}  *F -
iaF )  +
\tilde B \wedge (e^{+2\phi}  *\tilde G - ia\tilde G ) \biggr] \, .
\end{equation}

\noindent Thus, using Gauss's Theorem, we see that
\begin{equation}
I_{on-shell} = \int_{\partial M} \Xi \ ,
\end{equation}
where $\partial M$ is the boundary of $M$.

\subsection{Gibbons-Hawking type calculation of the Euclidean action}

To perform the explicit calculation of the Euclidean action for the
$U(1)^{2}$
black holes described in ref. \cite{US} we only have to evaluate eq.
(\ref{useful}). However, since in general the answer would be
infinite, we have
to
subtract the $K_{0}$ term described in the first section \cite{GH}.
For
convenience we switch from $\omega$ to $K$; the quantity to be
evaluated
is\footnote{There is no axion, and we have only one $F$ and one
$\tilde{G}$
field.} \begin{eqnarray}
&I_{\phantom{}_{E}}^{\phantom{}^\infty}  =
I_{\phantom{}_{E}}(gauge)+I_{\phantom{}_{E}}^{\phantom{}^\infty}([K])
=
\nonumber \\
 \nonumber \\
                       & =  \frac{1}{16\pi} \int d^4x\ \partial_\mu
\biggl[
\sqrt{g}\,
                                 [A_\nu ({\mbox{e}}^{-2\phi}
F^{\mu\nu})+
\tilde B_\nu
                                 ({\mbox{e}}^{+2\phi}  \tilde
G^{\mu\nu})]
\biggr] +
                                \frac{1}{8\pi}\int^{r\rightarrow
\infty} d^{3}x
                                \sqrt{h}(K-K_{0})
\end{eqnarray}
where the coordinates $x^{\mu}$, the metric $g_{\mu\nu}$ etc. are
now Euclidean objects, and the superscript $\infty$  means again that
the only
boundary of these spacetimes is at $r\rightarrow\infty$.  The reason
is the
same as in Schwarzschild case and was explained in the first section.

We start by calculating the extrinsic curvature term.
The dilaton black hole metric is given by \cite{US}
\begin{eqnarray}
ds^2       & = & e^{2U} dt^2 - e^{-2U} dr^2 - R^2 d\Omega^2 \,
,\label{dilmet}
\\
e^{2U}    & = & \frac{(r-r_+)(r-r_-)}{R^2}\, , \\
R^2        & = & r^2 - \Sigma^2 \, ,\\
r_{\pm} & = & M \pm r_0 \ .
\end{eqnarray}
For the metric (\ref{dilmet}), the radius of the local 2-sphere is
$R$, rather
than  $r$.  The metric may be rearranged as
\begin{equation}
ds^2 = e^{2U} dt^2 - e^{-2U} ({dr \over dR})^2 dR^2 - R^2 d\Omega^2,
\end{equation}
so that
\begin{equation}
K = {\frac{r  [(r-r_+)(r-r_-)]^{\frac{1}{2}} }{R^2}} \Biggl[
{\frac{(r-M)R}{r(r-r_+)(r-r_-)}} + {\frac{1}{R}} \Biggr]
\end{equation}
and
\begin{equation}
\frac{1}{8 \pi} \int^{r} d^{3}x (K - K_{0}) =-\frac{\pi}{\kappa}
\Biggl[  r - M + \frac{r(r-r_+)(r-r_-)}{R^2} - 2
[(r-r_+)(r-r_-)]^{\frac{1}{2}}
\Biggr] \ .
\end{equation}

Evaluating the limit of this expression at $r\rightarrow\infty$,
we get
\begin{equation}\label{KatInf}
I_{\phantom{}_{E}}^{\phantom{}^\infty}({K}) =  +\frac{\pi}{\kappa}
M\, .
\end{equation}

We now proceed to the evaluation of $I_{\phantom{}_{E}}(gauge)$ .
The gauge
and dilaton
fields are given by
\begin{eqnarray} F                & = & \frac{Q
e^{\phi_0}}{(r-\Sigma)^2} dt \wedge dr\ ,
                            \qquad A = \frac{Q
e^{\phi_0}}{(r-\Sigma)} dt\ ,
         \\
{\tilde{G}} & = &\frac{P e^{-\phi_0}}{(r+\Sigma)^2} dt \wedge dr\ ,
                           \qquad \tilde{B} = \frac{P
e^{-\phi_0}}{(r+\Sigma)}
dt\ ,  \\
e^{2\phi}   & = & e^{2\phi_0}\frac{r+\Sigma}{r-\Sigma}\ .
\end{eqnarray}

There is a subtlety involved in evaluation of the surface integral of
the
gauge terms.   Gibbons and Hawking argue in their treatment of the
Reissner-Nordstr\"{o}m black hole that, since the gauge potentials
are singular
on the event horizon $r_h= r_+$ (due to the vanishing of $g_{tt}$),
one must
make a gauge transformation to render them zero there,
\begin{equation}\label{Aprime}
A_\mu^{'}(r) = A_\mu(r) - A_\mu(r_h)  \,  .
\end{equation}
The gauge integrals become
\begin{equation}\label{gaugeterms}
 I_{\phantom{}_{E}}(gauge) = \lim_{r\rightarrow\infty} \Biggl[
\biggl[  Q^2
\biggl(
\frac{1}{r-\Sigma} -   \frac{1}{r_{+}-\Sigma} \biggr)+ P^2 \biggl(
\frac{1}{r+\Sigma} -   \frac{1}{r_{+}+\Sigma} \biggr)
\biggr]\int_{0}^{
\frac{2\pi}{\kappa}} \frac{d\tau}{2} \int \frac{d \Omega}{4\pi}
\Biggr]   \, .
 \end{equation}

In fact, we have found that the prescription (\ref{Aprime}) may
be thought of in another way, which ends up producing the same
result.

If we do perform the gauge transformation (\ref{Aprime}), it is clear
that
there
is no contribution from the horizon to  $I_{\phantom{}_{E}}(gauge)$,
since the
gauge-transformed vector potentials and therefore the integrand
vanish on it.
If, on the other hand, we do not wish to perform such a gauge
transformation,
then we must make a  careful consideration of Gauss's theorem. The
surface
integral has to be calculated on the boundary of the region in which
the
potentials are well-behaved and defined. In the case at hand, this
boundary
includes the horizon, and the integrand no longer vanishes there.
Nevertheless,
it turns out that the functional dependences of the gauge and dilaton
fields
conspire in such a way as to reproduce the previous result
(\ref{gaugeterms}).
To see this, consider the surface integral for the $F$ term:
\begin{eqnarray}\label{qrss}
-\frac{1}{8\pi} \int_{\partial M} d\sigma \,
n_\mu \biggl[ e^{-2\phi} F^{\mu\nu} A_\nu \biggr]
 &=& \frac{1}{8\pi} \int_{r_{+}}^{r\rightarrow\infty} dt  d\Omega
\Biggl[ R^2
e^{-2\phi_0} \, \biggl(\frac{r-\Sigma}{r+\Sigma}\biggr) \,  \frac{Q
e^{\phi_0}}{(r-\Sigma)^2}  \, \frac{Q e^{\phi_0}}{(r-\Sigma)} \Biggr]
\nonumber
\\
 \nonumber
\\
&=& \int_{r_{+}}^{r\rightarrow\infty} \frac{d\tau}{2}
\frac{d\Omega}{4\pi} \Biggl[ \frac{Q^2}{(r-\Sigma)} \Biggr] \ .
\end{eqnarray}
A similar thing happens for the $G$ term. Looking at eq.(\ref{qrss}),
we see  that the result for the  surface integral is simply
\begin{equation}\label{iegauge}
 I_{\phantom{}_{E}}(gauge) = -\frac{\pi}{\kappa} \Biggl[
\frac{Q^2}{(r_{+}-\Sigma)}  +
\frac{P^2}{(r_{+}+\Sigma)} \Biggr] \ ,
\end{equation}
which coincides with
eq.(\ref{gaugeterms}) obtained by doing the gauge transformation
demanded by
Gibbons and Hawking.

Finally, putting together eqs. (\ref{KatInf}), (\ref{iegauge}) for
the
extrinsic curvature term at infinity and the gauge terms, we get
\begin{equation}
I_{\phantom{}_{E}}^{\phantom{}^\infty} = \frac{\pi}{\kappa} \Biggl[ M
-
\frac{Q^2}{r_{+}-\Sigma} -
\frac{P^2}{r_{+}+\Sigma} \Biggr] \ , \label{mystery}
\end{equation}
which may be rearranged using the relations
\begin{equation}
\Sigma = \frac{P^2-Q^2}{2M} \, ,   \qquad
r_0^2 = M^2 + \Sigma^2 - P^2 - Q^2
\end{equation}
to give

\begin{equation}\label{Iinf}
I_{\phantom{}_{E}}^{\phantom{}^\infty} = \frac{\pi}{\kappa}r_0 =
\pi(r_+^2 -
\Sigma^2) =
\frac{1}{4} A(r_h) \, .
\end{equation}

For extreme dilaton black holes, this expression reduces to
\begin{equation}\label{extremeentropy}
I_{\phantom{}_{E}}^{\phantom{}^\infty}(extreme) = 2\pi |PQ| \, .
\end{equation}

Thus the method of Gibbons and Hawking, generalized to dilaton black
holes,
gives the result that the on-shell action coincides with the entropy
and is one
quarter of the area of the event horizon \footnote{For $PQ=0$, the
result of
the
calculation of the entropy was given  previously in \cite{HW}. It
agrees  (at
$a=1$)  with our expression (\ref{Iinf}) taken in the appropriate
limit .}.

 \section{On-shell Lagrangian for extreme $N=2$ black holes}

In section 3, the generalization of the Gibbons-Hawking method of
the calculation of the Euclidean action for dilaton black holes was
presented.
One starts with non-extreme black holes characterized by some finite
temperature
and surface gravity, and performs the calculation of the
on-shell action in Euclidean signature
by compactifying the Euclidean time coordinate. It turns out that one
can
express the total action as a surface
integral, and care has to
 be taken to
evaluate the contribution from the extrinsic curvature only from
spatial
infinity. As a final step, the extremal limit can be considered, and
the result
is:
\begin{equation}\label{A}
 S_{stringy} =  {\textstyle\frac{1}{4}} A = \pi (M^2 - \Sigma^2) =
{\textstyle\frac{1}{2}} \pi |z_1^2 - z_2^2|\ ,
\end{equation}
where $z_1$, $z_2$ are the central charges of extreme black holes
defined in \cite{US}.
This shows that when $N=2$ supersymmetry is restored, which takes
place when
the central charges are equal,
\begin{equation}
|z_1| = |z_2| \ ,
\end{equation}
the action vanishes. This was never possible for classical extreme
Reissner-Nordstr\"om black holes. Indeed those black holes are
solutions
of $N=2$ supergravity, which are characterized by a super-Poincar\'e
algebra at infinity with only one
central charge\footnote{The maximum number of central charges is
$N/2$ for
even $N$.}
\begin{equation}
 S_{RN} =  {\textstyle\frac{1}{4}} A = \pi M^2  =
{\textstyle\frac{1}{2}} \pi |z^2|\ .
\end{equation}
The stringy black holes are solutions of $N=4$ supergravity, and the
restoration of $N=2$ supersymmetry is the restoration of $O(2)$
internal
symmetry, which makes the two central charges equal and the action
(entropy)
vanish in agreement with the fact that the area of the horizon for
these
solutions
in the canonical geometry is zero.

Since the presence of the dilaton has  radically changed the
properties of
extreme
black holes, it becomes possible to address the following problem:
Could we
calculate the partition function for the extreme dilatonic black hole
{\it directly},
avoiding the intermediate step of introducing the concept of a
temperature at
all? The answer is positive for maximally supersymmetric purely
magnetic
(electric) extreme black holes, as we will now show.

Our starting point for the calculation of the on-shell action will be
the
Lagrangian in
 \begin{equation}\label{so4action}
 I =\frac{1}{16\pi} \int d^4x\,\sqrt{-g} \Biggl[ -R +
2\partial^\mu \phi \cdot \partial_\mu \phi- \left(
{\mbox{e}}^{-2\phi}F^{\mu\nu}F_{\mu\nu} + {\mbox{e}}^ {2\phi}\tilde
G^{\mu\nu}\tilde G_{\mu\nu}\right) \Biggr]\ ,
\end{equation}
with the additional $K$-term which removes the second derivatives
of the metric from the Lagrangian.
The gravitational part of the
action is given by eq. (\ref{LL}) and the vector $\omega^\mu$ in the
total
derivative
term in the  Lagrangian is given by eq.(\ref{eqforomega}).
Eq.(\ref{eqforomega}) can be also given in the form
\begin{equation}\label{landau}
\omega^\mu = - \frac{1}{\sqrt{-g}}\partial_\lambda  \left(
\sqrt{-g}g^{\lambda\mu}\right)  - g^{\lambda\mu}\left(
\partial_\lambda
\ln {\sqrt{-g}}
\right) \ .
\end{equation}
For our calculations there will be no need to rewrite the volume
integral for the  total derivative part in the Lagrangian (second
term in eq.
(\ref{LL})) as a  surface integral ($K$-term). Also we will not
transform the
gauge part of  the action to a surface integral as we did in the
previous
section.
It will be sufficient to keep all terms in a volume integral in what
follows.

The dilaton part of the Lagrangian is
\begin{equation}
\sqrt{-g}\,{\cal L}_{dil} =  \sqrt{-g}\,
2\partial^\mu \phi \cdot \partial_\mu \phi  \ .\end{equation}
The gauge part of the Lagrangian for the purely magnetic solution is
\begin{equation}
\sqrt{-g}\,{\cal L}_{ gauge} =
\sqrt{-g}\,{\cal L}_{ magn} = - \sqrt{-g}\,{\mbox{e}}^ {2\phi}\tilde
G^{\mu\nu}\tilde G_{\mu\nu}\ ,   \end{equation}
and for the purely electric
\begin{equation}
\sqrt{-g}\,{\cal L}_{ gauge} =
\sqrt{-g}\,{\cal L}_{ electr} = - \sqrt{-g}\,
{\mbox{e}}^{-2\phi}F^{\mu\nu}F_{\mu\nu}\ .
\end{equation}
The total action is given by the volume integral
\begin{equation}
16 \pi I = \int d^4x\, \sqrt{-g}\, \left( {\cal L}_{grav} +
{\cal L}_{dil} + {\cal L}_{ gauge} \right) \ ,
\end{equation}
The maximally supersymmetric purely magnetic (electric)
extreme black holes are  described by the following metric \cite{US}
\begin{equation}\label{metr}
ds^{2} =
{\mbox{e}}^{2U}dt^{2}-{\mbox{e}}^{-2U}d\vec{x}^{2}\ .
\end{equation}
Before using the field equations, let us calculate the total
derivative term
in the gravitational part of the Lagrangian for the ansatz
(\ref{metr}).
We find using eq. (\ref{landau}) that
\begin{equation}
\partial_\mu  \left(\sqrt{-g}\, \omega^\mu\right) = - 2 \, \partial_i
\partial_i
U\ . \end{equation}
The total gravitational part of the Lagrangian becomes
\begin{equation}
\sqrt{-g}\,{\cal L}_{grav} = \sqrt{-g}\,(- R)
 - 2 \, \partial_i \partial_i U \ .
 \end{equation}
At this stage we may start taking the equations of motion into
account. The dilaton for maximally supersymmetric extreme black holes
is related to the metric as follows:
\begin{equation}
\phi = \pm U \ ,
\end{equation}
where $(-)$ is for magnetic and $(+)$ for electric solution.
The first equation of motion which will be used to calculate the
on-shell
Lagrangian is the one which relates the scalar curvature to the
dilaton
contribution, see eq. (\ref{R}). It  follows that, on-shell,
\begin{equation}
\sqrt{-g}\, \left( {\cal L}_{grav} +
{\cal L}_{dil} \right) =
- 2 \, \partial_i \partial_i U\ .
\end{equation}
To treat the gauge action we have to use another equation of
motion,
\begin{equation}
\nabla^{2}\phi -
{\textstyle\frac{1}{2}} e^{-2\phi}F^{2} +
{\textstyle\frac{1}{2}} e^{2\phi}\tilde G^{2} = 0 \ .
\end{equation}
For the electric solution with $U= \phi$ it leads to
\begin{equation}
\sqrt{-g}\,{\cal L}_{ electr} = 2 \, \partial_i \partial_i \phi\ ,
 \end{equation}
and the total on-shell Lagrangian becomes
\begin{equation}
\sqrt{-g}\,{\cal L} = -2 \, \partial_i \partial_i U + 2 \, \partial_i
\partial_i \phi
=0 \ .
 \end{equation}
For the magnetic solution with $U= - \phi$
\begin{equation}
\sqrt{-g}\,{\cal L}_{ magn} = - 2\,  \partial_i \partial_i \phi
 \end{equation}
and the total on-shell Lagrangian becomes
\begin{equation}
\sqrt{-g}\,{\cal L} = -2 \, \partial_i \partial_i U - 2 \, \partial_i
\partial_i \phi
=0 \ .
\end{equation}
Thus the Lagrangian
and therefore the action vanishes for $PQ=0$
dilaton black holes, in agreement with eq. (\ref{Iinf}).  In other
words, we
get the
same result by evaluating the action directly as we get by taking the
$PQ=0$
limit of the expression for black holes with regular horizon.

Note that, in the process of calculation of the on-shell Lagrangian
for
maximally
supersymmetric extreme dilatonic black holes, we never faced the
problem of
going to Euclidean signature, choosing a proper gauge for the vector
potentials,
and thinking about boundary surfaces (the horizon versus infinity).
All of those problems were present in a standard treatment and, as
explained in
the previous section, can all be solved in quite satisfactory ways.
It is
therefore encouraging that an independent calculation of the action
exists, as given above for the extreme purely magnetic (electric)
black  holes that is consistent with the general formula that the
action is one
quarter of the area of the event  horizon.

\section{Discussion}

In this paper we have found that the entropy of general static
spherically symmetric black holes with a regular event horizon
 is given
by evaluating only the extrinsic
curvature term at the horizon and is one quarter of the area of the
event
horizon.
This generalizes the corresponding result derived for the
Schwarzschild black
hole by Hawking in \cite{Hawk100}.

For charged dilaton black holes, we have performed the calculation
both of the
action and of the entropy by following Gibbons and Hawking \cite{GH}.
In calculating the on-shell bosonic action for the theory in which
the dilaton
black hole is embedded, we have seen that it is topological and thus
may be
written as a surface integral.  We have found that the entropy
coincides with
the
on-shell action, in agreement with what one might expect from scaling
arguments as in \cite{Hawk100}.  Investigation of the action versus
entropy of
axion-dilaton black holes \cite{Tomas} is in progress.

A remaining puzzle is the physical origin of the entropy of $U(1)^2$
dilaton
black holes.  Extreme dilaton black holes, which could be the stable
endpoints of the evaporation process, may be thought of as
`groundstates'. In
the
theory we consider, the charges $P$ and $Q$ are central charges
(in different $U(1)$ groups), and there are no elementary charged
particles to
discharge the black hole.  These black holes also have zero
temperature and
unbroken $N=1$ supersymmetry \cite{US}, but the entropy is nonzero.
For these
extreme black holes, the entropy is given by  eq.
(\ref{extremeentropy}),
$S = 2\pi |PQ|$.   In \cite{US},
 we formulated a supersymmetric
nonrenormalization theorem which says that the result
(\ref{extremeentropy})
remains intact to higher order (perturbative) corrections in the
supersymmetric
theory.

In a quantum mechanical system,
entropy at zero temperature usually corresponds to degeneracy of the
groundstate. However, for the charged dilaton black holes the
relation between
the entropy and the degeneracy of these configurations is missing:
what kind of
``internal" degrees of freedom does the degeneracy correspond to?
Since the degeneracy of a quantum ground state is an integer, one may
then be
tempted to speculate\footnote{We thank L. Susskind for suggestions on
this
point.} that the entropy of extreme black holes is subject to the
quantization
rule
\begin{equation}\label{rule}  S = \frac{A}{4} = 2\pi |PQ| = \ln(n) \
,
\end{equation}
where $n$ is an integer and the area is measured in Planck units.
The size
of
the horizon is then not arbitrary, but restricted by the rule
(\ref{rule}).
Then purely magnetic or purely electric black holes have $n=1$ and
are clearly
allowed; they already have zero entropy and area.

Another possibility is to take seriously the fact that the quantity
$e^{A/4}$
is
generically not an integer, and that we  do not know  about the
existence of
any
internal degrees of freedom of the extreme electric-magnetic black
holes
responsible for the degeneracy of the state. Therefore, a possible
conclusion
is
that this state is {\it not}   a ground state of a quantum-mechanical
system,
having non-integer $e^ S$. Then we are led to a resolution of the
problem:
quantum-mechanically the extreme electric-magnetic black holes have
to be
unstable under
 splitting to another configuration of extreme black holes which
{\it is} a ground state and does have an
integer value $n=1$.

 The possibility that black holes may quantum mechanically split into
other
black holes was proposed in  \cite{US}.\footnote{Splitting of black
holes is
closely related to the possibility of splitting of the  universe into
many baby
universes. A particularly relevant example is splitting of one
Robertson-Bertotti universe into many RB universes, as discussed by
Brill
\cite{BrillRB}. For a recent discussion of splitting of dilaton black
holes
with
massive dilaton fields see  \cite{Hor}.}   A specific example
appropriate for
the issue of the ground state would be the splitting of the extreme
electric-magnetic black hole into a purely magnetic and a purely
electric one.
 Such bifurcation  is forbidden classically but  could in principle
occur in a
quantum-mechanical process and may be enforced by quantum-mechanical
instability of the zero temperature state with  non-integer value of
$e^S$ .
It can
be described by
\begin{equation}
(P, Q) \rightarrow (P,0) + (0, Q) \ .
\end{equation}

The extreme
electric-magnetic black hole has the following relations between
parameters:
\begin{equation}
M= \frac {|P|+ |Q|}{\sqrt{2}} \ ,  \hskip 2 cm \Sigma = \frac {|P|-
|Q|}{\sqrt{2}}\ ,
\end{equation}
which gives
\begin{equation}
M^2 + \Sigma^2 - P^2 - Q^2 = 0 \ .
\end{equation}
The parameters of the daughter black holes are related to those of
the parent
as
\begin{eqnarray}
\label{multi}
M&=& M_1 + M_2 \ ,\hskip 2 cm M_1 = \Sigma_1 = \frac
{|P|}{\sqrt{2}}\nonumber\\ \nonumber\\
\Sigma&=& \Sigma_1 + \Sigma_2 \ , \hskip 2 cm
M_2 = -\Sigma_2 = \frac {|Q|}{\sqrt{2}}\ .
\end{eqnarray}
After splitting the total entropy is equal to zero, and
\begin{eqnarray}
M_1^2 + \Sigma_1^2 - P^2  &=& 0 \ ,\nonumber\\
\nonumber\\
M_2^2 + \Sigma_2^2  - Q^2 &=& 0 \ .
\end{eqnarray}
These black holes are in an equilibrium with each other, since the
attractive
force between them vanishes due to supersymmetry \cite{US}.
Indeed, let us consider Newtonian,
Coulomb and
dilatonic forces.  The force between two distant objects of masses
and charges
$(M_1, Q_1, P_1, \Sigma_1)$ and $(M_2, Q_2, P_2,\Sigma_2)$ is
\begin{equation}\label{bal}
F_{12} = - \frac{M_1 M_2}{r_{12}^2} +
\frac{Q_1 Q_2}{r_{12}^2} +
\frac{P_1 P_2}{r_{12}^2} - \frac{\Sigma_1 \Sigma_2}{r_{12}^2} \ .
\end{equation}
The dilatonic force is attractive for charges of the same sign and
repulsive for charges of opposite sign.
Using the relations (\ref{multi}) for the masses and dilaton charges
 in terms of the magnetic and electric charges $P_1 = P$, $P_2 = 0$,
$Q_1 = 0$,
$Q_2 = Q$, we
see  that $F_{12}$ vanishes.

 Thus, after splitting a possible ground state of
the quantum mechanical system is reached  which could be a pure state
with
$S = 0$.\footnote{Another example of a solution with zero entropy, is
the
supersymmetric domain wall \cite{MIR}.  The authors argue that it
describes a
non-degenerate groundstate.} The area of the horizon of both black
holes is now
zero.
 Equation (\ref{multi}) describes the distribution of masses and
charges in  a
particular example of
the general extreme supersymmetric multi-black hole  solution, given
in
\cite{US}. Will the purely electric and purely magnetic black holes
continue
 splitting
to the smallest values of charges? Is the value of the entropy
$S=2\pi
|PQ|$ responsible for the degeneracy properties of the ground state?
These and
many other questions can be asked in connection with the calculated
value of
the
entropy of charged dilatonic black holes.

\section*{Acknowledgements}
The authors wish to thank D. Brill, G. Horowitz, M. Perry,  A.
Strominger, L. Thorlacius and K. Thorne for useful discussions.
Our interpretation of the possible physical meaning of the entropy
of the extreme dilaton black holes originated from numerous
discussions with
A. Linde and L. Susskind. One
of us (R.K.) would like  to thank the Aspen Center for Physics and
participants
of
the  Black Hole workshop for stimulating conversations.

The work of R.K. and A.P.  was supported by NSF grant PHY-8612280.
The work
of R.K.  was supported in part by the John and Claire Radway
Fellowship in the
School of Humanities and Sciences at  Stanford University.  The work
of T.O.
was
supported  by a Spanish Government M.E.C. postdoctoral grant.

\end{document}